\documentclass[conference]{IEEEtran}
\IEEEoverridecommandlockouts
\usepackage{cite}
\usepackage{amsmath,amssymb,amsfonts}
\usepackage{algorithmic}
\usepackage{graphicx}
\usepackage{textcomp}
\usepackage{xcolor}
\usepackage[colorlinks=true,linkcolor=black,anchorcolor=black,citecolor=black,filecolor=black,menucolor=black,runcolor=black,urlcolor=black]{hyperref}
\def\BibTeX{{\rm B\kern-.05em{\sc i\kern-.025em b}\kern-.08em
    T\kern-.1667em\lower.7ex\hbox{E}\kern-.125emX}}
\begin{document}

\title{Understanding the Communist Party of China's Information Operations
}

\author{\IEEEauthorblockN{Rohit Dube}
\IEEEauthorblockA{\textit{Independent Researcher} \\
California, USA \\
}
}

\maketitle

\begin{abstract}
The Communist Party of China is known to engage in Information Operations to influence public opinion.
In this paper, we seek to understand the tactics used by the Communist Party in a recent Information Operation - the one conducted to influence the narrative around the pro-democracy movement in Hong Kong.
We use a Twitter dataset containing account information and tweets for the operation.
Our research shows that the Hong Kong operation was (at least) partially conducted manually by humans rather than entirely by automated bots.
We also show that the Communist Party mixed in personal attacks on Chinese dissidents and messages on COVID-19 with the party's views on the protests during the operation.
Finally, we conclude that the Information Operation network in the Twitter dataset was set up to amplify content generated elsewhere rather than to influence the narrative with original content.
\end{abstract}

\begin{IEEEkeywords}
Information Operations, Data Science, Communist Party of China, Hong Kong, Twitter, COVID-19
\end{IEEEkeywords}

\section{Introduction}
The Communist Party of China (CCP) engages in Information Operations (IO) in foreign countries to sow disinformation and misinformation \cite {hks1}. 
The CCP's IO tactics have been studied by a handful of researchers \cite{beskow1}, \cite{ferrara1}, \cite{nimmo2}, \cite{kao1}.
However, to the best of our knowledge, a June 2020 CCP IO dataset from Twitter \cite {twitter1} \cite{twitter2} has not been studied by researchers.
Our goal is to examine this Twitter dataset to learn about the structure of China's IO operations.

\section{Data Background}
In May 2020, Twitter deleted more than $23K$ accounts linked to the CCP \cite{twitter1}.
These accounts were responsible for $348K+$ tweets on Hong Kong's pro-democracy protests, Chinese dissidents and COVID-19.
Twitter officially attributed its action to ``... CCP ... continuing to push deceptive narratives about the political dynamics in Hong Kong.''
Shortly after deleting the CCP controlled accounts, Twitter made information on the deleted accounts public \cite{twitter2}.
This Twitter dataset gives us a window into the CCP's IO infrastructure.

Twitter is blocked in mainland China but not in Hong Kong \cite{twitter3}. 
Users in mainland China, including those working for the CCP, can access Twitter by using a virtual private network (VPN) service.
Further, the CCP's IO network sent messages in mixed Chinese and English. 
Given the difficulty of accessing Twitter in mainland China and the languages used in the tweets, we believe that the network's primary target was the citizenry of Hong Kong and Chinese nationals living outside China (and not the people of mainland China).

We use two Comma-separated Values (CSV) files from the Twitter dataset.
The first file (china\_052020\_users\_csv\_hashed.csv) contains information on the accounts deleted by Twitter.
Twitter recorded and made public the account creation date, follower count (number of accounts following an account) and following count (number of other accounts followed by an account) among other information, for each deleted account.

The second CSV file (china\_052020\_tweets\_csv\_hashed.csv) contains information on each tweet sent out by the deleted accounts.
Twitter makes available the account that sent out the tweet, along with other information on the tweet such as tweet time, hashtags used and other Twitter accounts mentioned by the tweet.
Twitter also records the number of times each tweet was quoted, replied to, liked or retweeted.

\section{Related Work}
\cite{jankowicz1} lays out the CCP's rationale for engaging in IO campaigns as part of ``public opinion or media warfare.'' 
\cite{nimmo1} provides a yardstick to measure the impact of IO campaigns.
\cite{milanovic1} attempts to understand the obligations of states concerning IO campaigns as per existing international law.
While these papers primarily focus on IO in a country foreign to the attacker, they also provide a backdrop for our research.

\cite{beskow1} uses data science techniques to study $206$ million tweets on COVID-19 from March 15 to April 30, 2020.
This study finds that CCP COVID-19 IO operations were primarily focused on the United States, were more aggressive than past CCP IO operations, and used human operators in addition to bots to shape Hong Kong and COVID-19 discourse on Twitter.
\cite{ferrara1} studies $43.5$ million tweets from the early days of the COVID-19 pandemic and documents the use of bots in promoting political conspiracies.
Both studies are instructive, but neither uses the Twitter dataset (published after these studies went to press.)

\cite{nimmo2} identified a (presumed) CCP IO network active across multiple social media platforms and dubbed the ``Spamouflage Dragon.'' 
This IO network has been active at least since September 2019. 
The study documents the types of social media posts used by the Spamouflage Dragon.
\cite{kao1} is similar and documents the CCP's efforts to commandeer existing social media accounts for use in IO campaigns.
Our research extends these studies -- it seeks to add to the literature on the CCP's IO tactics.

\section {Analysis}
We analyzed the Twitter dataset using standard  data science techniques, including top-K ranking by attribute, topic modeling, and network centrality measurements.

\subsection {Account creation}
Pro-democracy protests in Hong Kong escalated in October 2019 \cite{reuters1}.
There is evidence that the escalation was the motivating force behind the IO network and campaign.

\begin{figure}
	\includegraphics[scale=0.25]{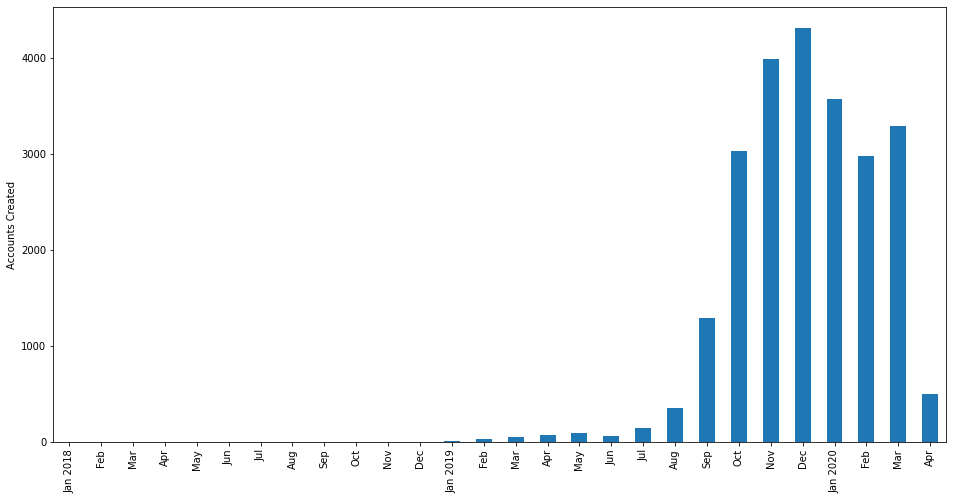}
	\caption {\label{fig: account-creation} New account creation by year / month}
\end{figure}

Figure \ref{fig: account-creation} shows the account creation trend for the accounts in the Twitter dataset. 
Most of the CCP accounts deleted in May 2020 were created between October 2019 and March 2020.

\begin{figure}
	\includegraphics[scale=0.25]{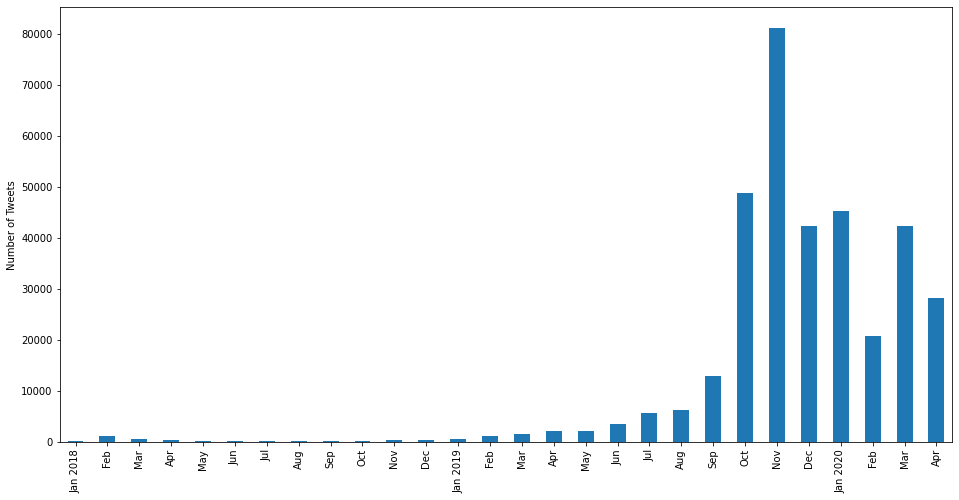}
	\caption {\label{fig: tweets-by-date} Tweets by year / month}
\end{figure}

Figure \ref{fig: tweets-by-date} shows the volume of tweets sent out by the accounts in the dataset.
The volume of tweets roughly tracks the volume of account creation.
Note the jump in the volume for both metrics in October 2019.

\subsection {Tweet timing}
From the pattern of tweet timing, we conclude that the IO network was (at least) partially controlled by human operators who treated their work as a ``job.''

\begin{figure}
	\includegraphics[scale=0.25]{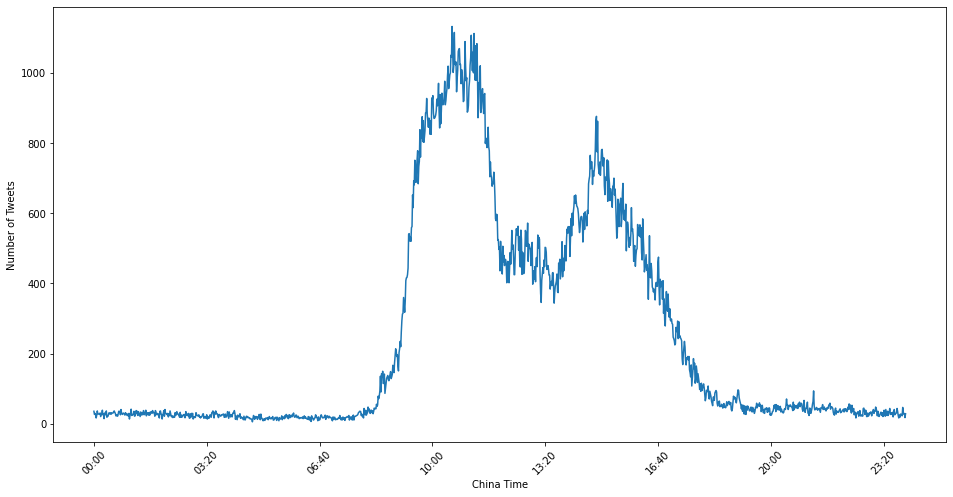}
	\caption {\label{fig: tweets-by-time} Tweets by time-of-day}
\end{figure}

Figure \ref{fig: tweets-by-time} plots the average number of tweets by the time of day (in local China time).
The intensity of tweets follows a typical workday schedule, including a lunch break.

\begin{figure}
	\includegraphics[scale=0.25]{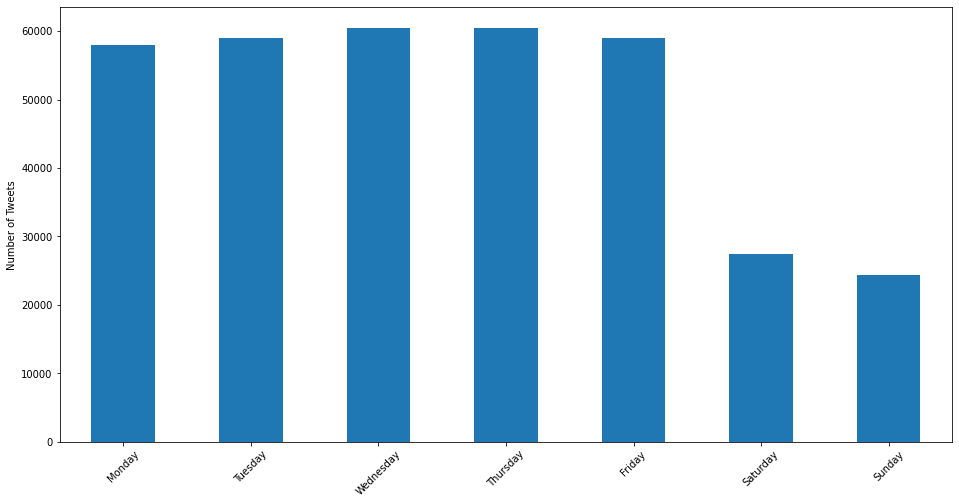}
	\caption {\label{fig: tweets-by-dow} Tweets by day-of-week}
\end{figure}

Figure \ref{fig: tweets-by-dow} plots the total number of tweets by the day of the week.
Many more tweets were sent out on weekdays compared to weekends, indicating that the network operators did less work over weekends compared to weekdays.

While it is possible for an IO network to be programmed to behave like one powered by humans, prior research \cite{berkeley1} suggests the involvement of CCP directed humans in IO campaigns.

\subsection {Top themes}
We used three methods to understand the main themes that the IO network was instructed to promote: top tweets, top hashtags, and topic modeling based on the tweet text.
Using these three methods, we find three clear areas of commentary from the IO network: the protest narrative in Hong Kong, Chinese dissidents, and the Covid-19 pandemic (the early stages coincident with the Hong Kong protests).

\textbf {Top tweets:} The five most popular tweets are identified by adding up the ``quote,'' ``reply,'' ``like,'' and ``retweet'' counts for each tweet from the IO network. 
The top-5 tweets are listed below:

\begin {enumerate}
        \item 
                \#HongKong Fully support the Hong Kong Police Force 
                to enforce laws strictly, stop riots and control chaos, 
                maintain Hong Kong's safety and stability, and revive 
                the glory of the past \#HongKong https://t.co/Cxjy3HljIO
        \item
                \#HongKong Commemorate the anniversary of the promulgation of the Hong Kong Basic Law https://t.co/mThf0SV1qR   
        \item
                \#GuoWengui life is becoming more and more difficult. The legal fund has changed to cheat and cheat. \#GuoWengui will end up with nothing. https://t.co/NiOjQOqkCF      
        \item
                \#HongKong Hong Kong, once an international financial center, has a vibrant and good atmosphere. It has a spiritual core of "freedom, equality, \#HongKong and the rule of law", incites violence and discrimination, and affects the freedom of others. \#HongKong \#virus https://t.co/prriQf2x40     
        \item
                \#Covid19 \#USAVirus   Viruses have no borders. It's a global battle. No one can survive. No country can survive alone. \#CoronavirusDisease   https://t.co/QoB25wSucI  
\end {enumerate}

Chinese words (tokens) in the tweets were translated into English using Google Translate \cite{google1}. 
The translation and the prevalence of mixed language (Chinese and English) sentences are partially responsible for the syntactically incorrect English in the tweets listed above.
Note that Google Translate was also used in other parts of this analysis to translate Chinese text into English.

The top-5 tweets promote ``law and order" in Hong Kong, denigrate a Chinese dissident (Guo Wengui) and promote CCP messaging on COVID-19 (that the pandemic is a global problem).

\begin{figure}
        \includegraphics[scale=0.22]{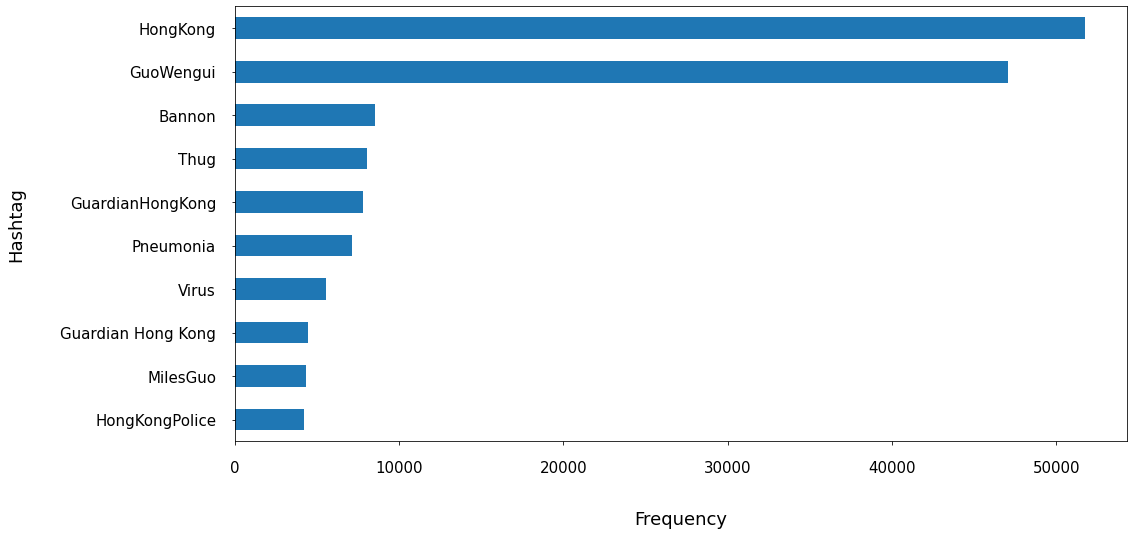}
        \caption {\label{fig: hashtags} Top 10 hashtags}
\end{figure}

\textbf {Top hashtags:} We also identified the top-10 hashtags used by the IO network. Figure \ref{fig: hashtags} lists these hashtags.
The hashtags corroborate the CCP's focus on law and order (\#GuardianHongKong), Chinese dissidents (\#Guo Wengui) and Covid-19 (\#Virus).

\textbf {Top topics:} Finally, tables \ref{tab: topic1}, \ref{tab: topic2}, \ref{tab: topic3}, \ref{tab: topic4}, \ref{tab: topic5} list the top-5 topics in the IO network's tweets.
We report the top-5 tokens (by weight/probability-of-occurrence) for each topic.

The token ``rt'' is an abbreviation for ``retweet.''
The token ``liyiping1911'' refers to a Chinese dissident in Canada - Li Yiping. 
Other tokens refer to another Chinese dissident - Guo Wengui - including by other names such as ``Miles Guo.''
``color revolution'' refers to anti-communism demonstrations in communist or formerly-communist countries \cite{wiki1}.
The CCP (incorrectly) attributes such color revolutions to incitement by the US and other democratic countries.

The top-5 topics confirm the observations from the top-5 tweets and top-10 hasthags regarding the focus of the IO network.

The topics and the tokens were identified using Non-negative Matrix Factorization (NMF) with Kullback-Leibler (KL) divergence \cite{nature1}.
Topic modeling using NMF was implemented using Python's Scikit-learn library \cite{python1}.
We restricted the number of features (tokens) used by the algorithm to $1000$ and removed English stop-words (such as ``a,'' ``and,'' ``the'') from consideration.
The algorithm was run multiple times with different ``random seeds.'' 
The results from the different runs were similar (but not always the same).

While the text of many of the tweets in the dataset included both English and Chinese, we believe that the topics found by NMF are valid.
This is because the implementation described above extracts each token separated by white space irrespective of the language. 
Subsequently, ``term frequency -- inverse document frequency (TFIDF)" is used to create a language-independent vector representation of each tweet.
The NMF algorithm uses these vector representations to determine the topics - i.e., by the time NMF is run, the tweet language has been abstracted away.

\begin{table}[htbp]
        \begin{center}
                \begin{tabular}{|c|c|}
                \hline
                \textbf{Token} & \textbf{Weight} \\ \hline
                rt & 10.799 \\ \hline
                guard hong kong & 0.416 \\ \hline 
                liyiping1911 & 0.344  \\ \hline
                parade & 0.304  \\ \hline
                mob & 0.279  \\ \hline
                \end{tabular}
        \end{center}
\caption {\label{tab: topic1} Topic 1: Law-and-order and Chinese dissident Li Yiping}
\end{table}

\begin{table}[htbp]
        \begin{center}
                \begin{tabular}{|c|c|}
                \hline
                \textbf{Token} & \textbf{Weight} \\ \hline
                guo wengui & 8.770 \\ \hline
                miles guo & 1.031 \\ \hline 
                breaking news revolution & 0.639  \\ \hline
                rule of law fund & 0.511  \\ \hline
                legal fund & 0.459  \\ \hline
                \end{tabular}
        \end{center}
\caption {\label{tab: topic2} Topic 2: Law-and-order and Chinese dissident Guo Wengui}
\end{table}

\begin{table}[htbp]
        \begin{center}
                \begin{tabular}{|c|c|}
                \hline
                \textbf{Token} & \textbf{Weight} \\ \hline
                hong kong & 8.309 \\ \hline
                thugs & 0.802 \\ \hline 
                guard hong kong & 0.662  \\ \hline
                stop violence and control chaos & 0.610  \\ \hline
                color revolution & 0.585  \\ \hline
                \end{tabular}
        \end{center}
\caption {\label{tab: topic3} Topic 3: Law-and-order and color-revolution}
\end{table}

\begin{table}[htbp]
        \begin{center}
                \begin{tabular}{|c|c|}
                \hline
                \textbf{Token} & \textbf{Weight} \\ \hline
                bannong & 5.622 \\ \hline
                wengui & 1.043 \\ \hline 
                new york times & 0.732  \\ \hline
                wall street journal & 0.577  \\ \hline
                nowadays & 0.544  \\ \hline
                \end{tabular}
        \end{center}
\caption {\label{tab: topic4} Topic 4: Chinese dissident Guo Wengui and western-media}
\end{table}

\begin{table}[htbp]
        \begin{center}
                \begin{tabular}{|c|c|}
                \hline
                \textbf{Token} & \textbf{Weight} \\ \hline
                pneumonia & 1.549 \\ \hline
                virus & 1.282 \\ \hline 
                epidemic & 1.199  \\ \hline
                covid19 & 0.999  \\ \hline
                youtube & 0.929  \\ \hline
                \end{tabular}
        \end{center}
\caption {\label{tab: topic5} Topic 5: Covid-19}
\end{table}

\subsection {Role of network}
The IO network was likely created to amplify CCP messaging.
We have a strong indication of the amplification role from retweets as a percentage of the total tweets sent by the network: $52\%$ of the tweets were retweets.

\begin{figure}
	\includegraphics[scale=0.25]{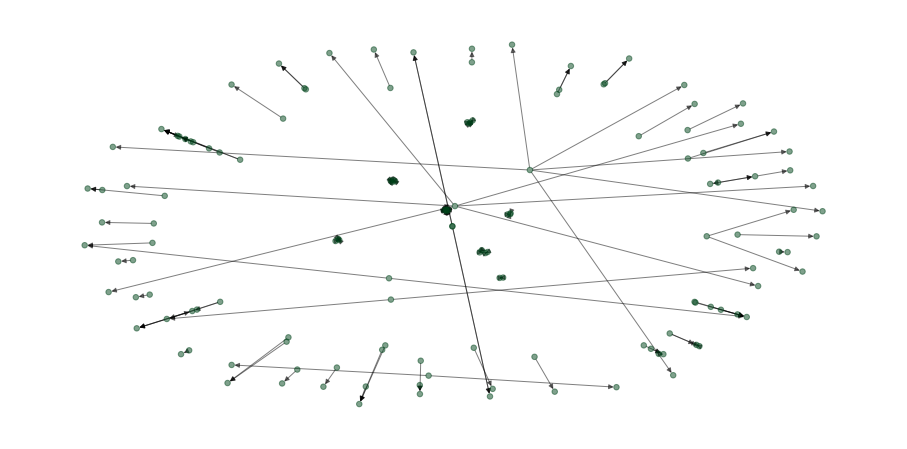}
	\caption {\label{fig: included-mentions} Mentions of accounts in the dataset by other accounts in the dataset (only mentions $\ge 20$ shown)}
\end{figure}

A network analysis of ``user mentions'' supports this conclusion.
Figure \ref{fig: included-mentions} shows a graph of nodes (accounts) mentioning other accounts in tweets when only the IO network taken down by Twitter is considered. 
We have removed links with user mention counts less than $20$ and all isolated nodes. 
Note that the links in the graph are directed - from the node representing the ``mentioner'' account to the node representing the ``mentioned'' account.
The nodes in figure \ref{fig: included-mentions} are shaded green, indicating that these nodes represent accounts that were part of the IO network taken down by Twitter.

\begin{figure}
	\includegraphics[scale=0.25]{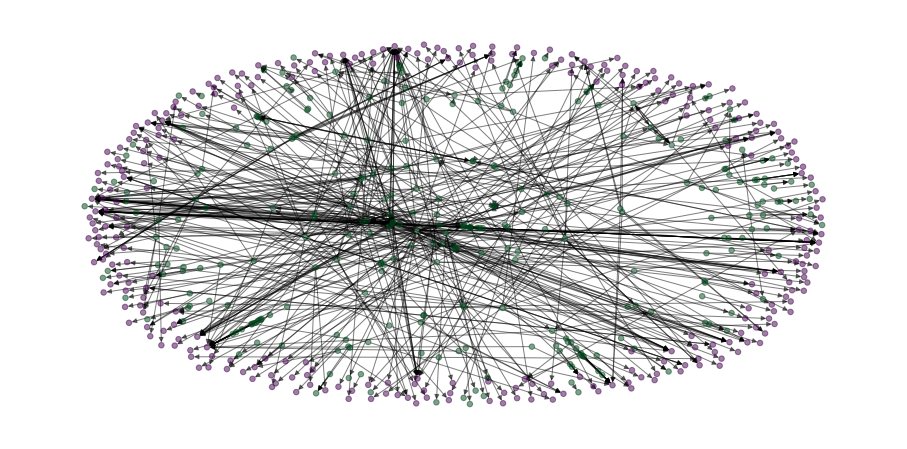}
	\caption {\label{fig: mixed-mentions} Mentions of all accounts by accounts in the dataset (only mentions $\ge 20$ shown)}
\end{figure}

Figure \ref{fig: mixed-mentions} shows the same graph but includes accounts that were not in the IO network.
That is, Figure \ref{fig: mixed-mentions} includes accounts outside the IO network that were mentioned by accounts in the IO network.
These ``external'' accounts are shaded purple for clarity.
Figure \ref{fig: mixed-mentions} shows these purple nodes on the outer edges of the graph.
In contrast, the green nodes are mostly on the inside of the graph.
Figure \ref{fig: mixed-mentions} is significantly more dense than figure \ref{fig: included-mentions} indicating that the IO network (green nodes) played an amplification role rather than a ``content creator'' role -- 
i.e., the accounts represented by the green nodes mentioned the accounts represented by the purple nodes much more often than they mentioned other green nodes.

We also extract the top-25 accounts by in-degree \cite{python2} - i.e., the 25 accounts that were mentioned most often.
We find that only three accounts in the top-25 were part of the network taken down by Twitter (i.e., three of the top-25 were green nodes and the rest purple).
This finding corroborates the role of the network as an amplifier that drives up popularity measures for tweets created outside the network.

\begin{figure}
	\includegraphics[scale=0.33]{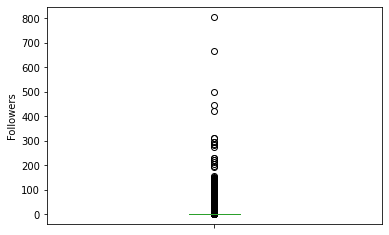}
	\caption {\label{fig: follower} Boxplot for number of followers}
\end{figure}

\begin{figure}
	\includegraphics[scale=0.33]{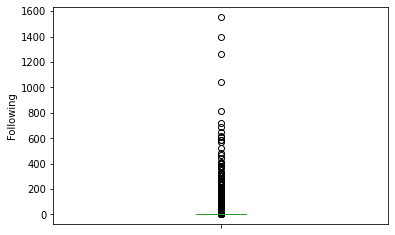}
	\caption {\label{fig: following} Boxplot for number of accounts followed}
\end{figure}

Finally, additional evidence for the amplification role of the network comes from the follower (figure \ref{fig: follower}) and following (figure \ref{fig: following}) counts of the green nodes.
Both counts are modest on average.
Even the accounts with high follower counts do not have enough followers to be considered influencers.

\section {Summary}
We believe that this is the first paper in the public domain to analyze Twitters June 2020 dataset \cite{twitter1}, \cite{twitter2}.

We have shown in this paper that the CCP responded to the pro-democracy protests in Hong Kong, in part, by setting up an IO network on Twitter.
This Twitter IO network was managed manually (at least partially). 
The network pushed the CCP narrative on the Hong Kong protests while attacking Chinese dissidents (such as Guo Wengui and Li Yiping) and promoting the CCP's messages on COVID-19.
Finally, the IO network was used to amplify content produced elsewhere rather than influence the narrative with original content.

\section*{Acknowledgment}
We want to thank reviewers of the research proposal that led to this work.
We would also like to thank those who reviewed the early drafts of this paper.

\end{document}